\documentclass[english,prl,aps,floatfix,superscriptaddress,twocolumn,longbibliography]{revtex4-1}
\usepackage[T1]{fontenc}
\usepackage[latin9]{inputenc}
\usepackage{color}
\usepackage{float}
\usepackage{bm}
\usepackage{amsmath}
\usepackage{amssymb}
\usepackage{graphicx}
\usepackage{wasysym}
\usepackage{cancel}
\usepackage[colorlinks,linkcolor=blue,anchorcolor=blue,citecolor=blue,urlcolor=blue]{hyperref}


\usepackage{babel}
\begin{document}
\renewcommand{\figurename}{FIG.}
\title{Perfect Crossed Andreev Reflection in Dirac Hybrid Junctions in the
Quantum Hall Regime}
\author{Song-Bo Zhang}
\affiliation{Institute for Theoretical Physics and Astrophysics$\text{,}$
 University of W\"urzburg, D-97074 W\"urzburg, Germany}
% \affiliation{W\"urzburg-Dresden Cluster of Excellence ct.qmat}
\author{Bj\"orn Trauzettel}
\affiliation{Institute for Theoretical Physics and Astrophysics$\text{,}$
 University of W\"urzburg, D-97074 W\"urzburg, Germany}
 \affiliation{W\"urzburg-Dresden Cluster of Excellence ct.qmat, Germany}
\date{\today}
\begin{abstract}
Perfect crossed Andreev reflection (CAR) is striking for high-efficiency
Cooper pair splitting which bears promising applications in quantum
communication. Recent experimental advances have disclosed the way
to explore CAR in Dirac fermion systems under ultra-strong magnetic
fields. We develop a scattering approach to study quantum Hall-superconductor-quantum
Hall (QH-S-QH) junctions formed by a two-dimensional (2D) time-reversal
symmetric Dirac semimetal. We propose two different  setups of the
hybrid junction in the quantum limit where only zeroth Landau levels
are involved in transport to exploit perfect CAR. In both setups,
the CAR probability can reach unity without applying bias voltage
and is controllable by the magnetic field strength, junction width,
length and doping of the superconductor. CAR dominates the nonlocal
transport and is directly measurable by the differential conductances.
We also identify a quantized spin injection per CAR event in one of
the two setups. Our proposal is experimentally feasible and will
be helpful for exploring high-efficiency Cooper pair splitting and
spin injection in Dirac materials.
\end{abstract}
\maketitle
\textit{Introduction.}---Crossed Andreev reflection
(CAR) is a process of converting an electron/hole from one lead to
a hole/electron in another lead through a superconductor (S) \cite{Byers95PRL,Recher01PRB,Lesovik01EPJB}.
Via CAR, Cooper pairs, which are strongly entangled electron pairs, can in principle be split spatially \cite{Recher01PRB,Lesovik01EPJB,Samuelsson03PRL}
and find fundamental interest and promising applications in quantum
communication \cite{Burkard00PRB,Horodecki09RMP,Modi12RMP}. Thus,
searching for systems with a large probability and convenient manipulation
of CAR is desirable. A variety
of candidate systems for CAR have been proposed, which include ferromagnetic junctions \cite{Deutscher00APL,Beckmann04PRL,Linder09PRB,KLi16PRB,Beiranvand17PRB},
p-n junctions \cite{Cayssol08PRL,Veldhorst10PRL}, topological systems
\cite{Nilsson08PRL,WChen11PRB,WChen12PRL,Reinthaler13PRL,JHe14ncomm,JWang15PRB},
and other platforms \cite{Russo05PRL,Hofstetter09nature,Cadden09nphys,JWei10nphys,Herrmann10PRL,Schindele12PRL, Das12ncommuns,ZBTan15PRL,Fulop15PRL,ZHou16PRB,Ang16PRB,Beconcini18PRB,Finocchiaro18PRL}.
Some of them have been experimentally implemented \cite{Beckmann04PRL,Russo05PRL,Hofstetter09nature,Cadden09nphys, JWei10nphys,Herrmann10PRL,Schindele12PRL,Das12ncommuns,ZBTan15PRL,Fulop15PRL}. Nevertheless, most proposals require a fine tuning of electronic structure or a
particular bias voltage. Usually, the processes of electron co-tunneling (EC)
and local Andreev reflection (LAR) are inevitable, which tend
to suppress and obscure CAR. It remains a challenge to have a system free of both detrimental processes.

The quantum Hall (QH) effect forces charged carriers to move along
chiral edge channels which are robust against disorder \cite{Laughlin81prb,Halperin82prb}.
Recently, hybrid systems cooperating with the QH effect and superconductivity
have been fabricated based on graphene \cite{Rickhaus12NL,Amet16science,GHlee17nphys,Park17Srep,Sahu18PRL,Seredinski19arXiv}
 whose low-energy physics is governed by Dirac fermions. This paves
a new way to explore CAR in Dirac materials. However, many physical
properties of Dirac hybrid structures in the QH regime, particularly
in the quantum limit where only lowest Landau levels contribute to
transport, have yet to be explored.

\begin{widetext}

\begin{figure}[th]
\centering

\includegraphics[width=16cm]{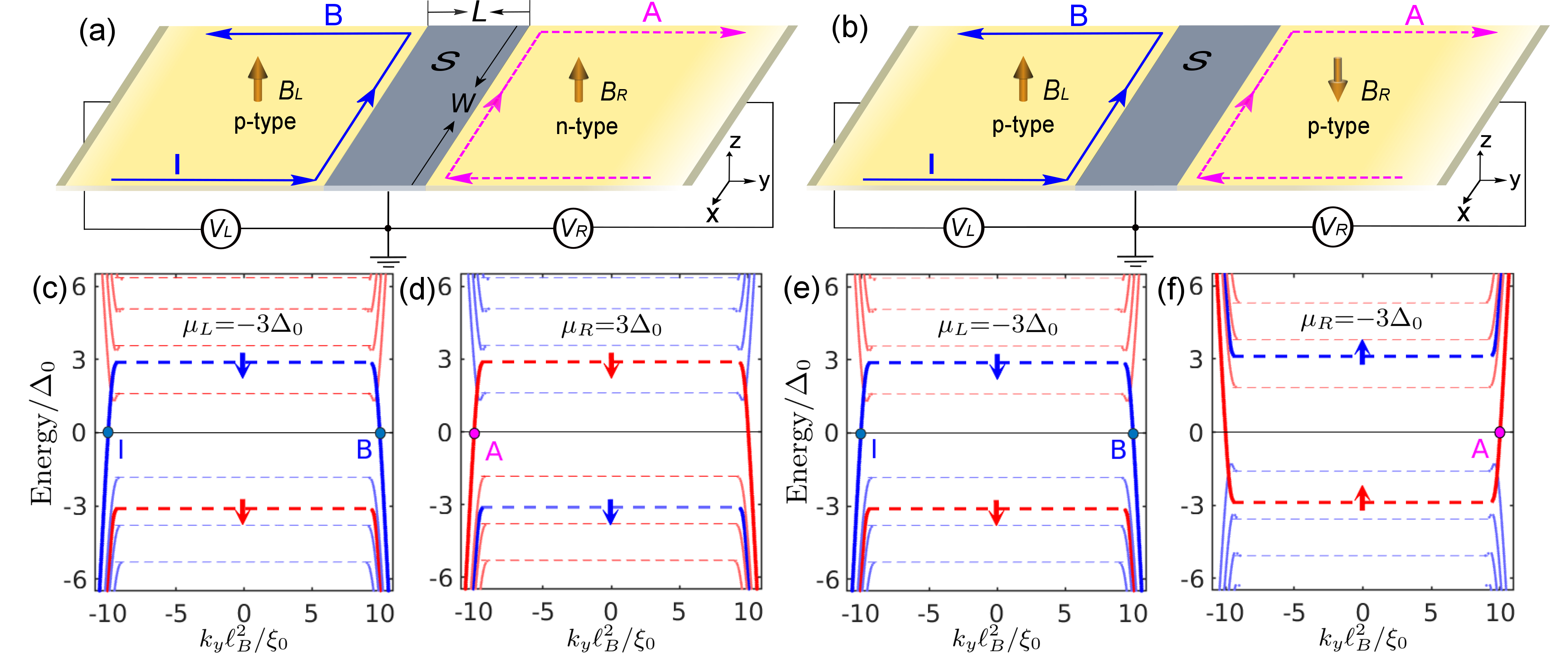}

\caption{(a) the p-S-n junction with $\mu_L\mu_R<0$ and $B_LB_R>0$.
``n-type'' and ``p-type'' refer to electron and hole doping, respectively.
Excitation energy spectra in the (c) left and (d) right QH regions. The red and blue lines are for electrons and
holes, respectively. The thick lines are zeroth Landau levels. The
arrows $\uparrow$($\downarrow$) indicate spin-up(-down) polarization. The dashed lines represent the bulk Landau levels given by Eqs.\ (\ref{eq:LandauLevel}). The locations
of the incident electron (I), normal reflected electron (B) and crossed-Andreev-reflected
hole (A) are indicated in (a). Here, $\mu_{R}$$=$$-\mu_{L}$$=$$3\Delta_0$,
$B_{L}$$=$$B_{R}$$=$$11B_0$, $\kappa=0.01v\xi_0$ and $W=20\xi_0$. (b) the n-S-n junction
with  $\mu_L\mu_R>0$ and $B_LB_R<0$. (e) and (f) are the same as (c) and (d) but for the n-S-n junction with $\mu_{R}$$=$$\mu_{L}$$=$$-3\Delta_0$ and $B_{L}$$=$$-B_{R}$$=$$11B_0$.}

\label{fig:schematic-nSp}
\end{figure}
\end{widetext}

In this Letter, we develop a scattering approach to investigate superconducting
hybrid junctions in the QH regime, which are based on 2D time-reversal
symmetric Dirac semimetals. In the quantum limit, the transport of
the Dirac semimetal is governed by particular zeroth Landau levels
which are spin polarized and chiral. Making use of this mechanism, we propose the QH-S-QH junction in two different setups of the quantum
limit as a novel platform for perfect CAR. One setup is a p-S-n junction with the same magnetic field but different types
of doping in the two QH regions, while the other one is an n-S-n junction
with opposite magnetic fields but the same type of doping, as sketched in Figs.\ \ref{fig:schematic-nSp}(a) and (b), respectively. Remarkably, in both setups, EC and LAR are completely blocked, and CAR can be enhanced without fine tuning of bias voltage. The CAR probability can reach unity and is influenced by the length and doping of the superconductor, magnetic field strength and junction width.
Due to the particular properties of conducting channels, CAR dominates the nonlocal transport and can be directly measured
by differential conductances. Moreover, we find that while there
is no spin injection in the n-S-n junction, a quantized spin injection
per CAR event occurs in the p-S-n junction, which suggests a new route
for high-efficiency spin injection in superconducting spintronics.

\textit{QH-S-QH junction based on a 2D Dirac semimetal.}---We
start with a time-reversal symmetric Dirac semimetal in two dimensions,
which is described at low energies by
\begin{equation}
H_{0}(\hat{{\bf k}})=\begin{pmatrix}H(\hat{{\bf k}}) & 0\\
0 & \mathcal{T}H(\hat{{\bf k}})\mathcal{T}^{-1}
\end{pmatrix}.\label{eq:originalModel}
\end{equation}
The basis function is $(\psi_{+,\uparrow},\psi_{+,\downarrow},\psi_{-,\uparrow},\psi_{-,\downarrow})$
with the indices $\pm$ labeling the two Dirac cones related by time-reversal
symmetry and $\uparrow$,$\downarrow$ the two spins. The effective Hamiltonian
reads $H(\hat{{\bf k}})=v\hat{k}_{x}s_{x}+v\hat{k}_{y}s_{y}+\kappa\hat{{\bf k}}^{2}s_{z}$
with $v$ the Fermi velocity, $\hat{{\bf k}}=(\hat{k}_{x},\hat{k}_{y})\equiv-i(\partial_{x},\partial_{y})$
the wavevector operators and $(s_{x},s_{y},s_{z})$ the
Pauli matrices acting on spin space. A small quadratic correction
$\kappa|{\bf k}|\ll v$ is introduced to regulate the topological
properties as ${\bf k}\rightarrow\infty$ and ensure definite edge
states \cite{Shen12book}. $\mathcal{T}=is_{y}\mathcal{C}$ is the
time-reversal operator with $\mathcal{C}$ the complex conjugation.
The model\ (\ref{eq:originalModel}) can be used to describe surface
states of 3D topological insulators \cite{HJZhang09nphys,YLChen09science,Lu10prb,Shan11njp}, and
the transition phase between a quantum spin Hall insulator and a normal
insulator \cite{Bernevig06science,Murakami07PRB,Buttner11nphys}.

The QH-S-QH junction under study is formed by the Dirac semimetal in a strip geometry, as depicted in Figs.\ \ref{fig:schematic-nSp}(a,b).
Without loss of generality, we take the junction in $\hat{y}$ direction
and apply the magnetic field $B_{L/R}$ in $\hat{z}$ direction in
the left/right normal-metal lead. The junction (or strip) width is
$W$ and the length of the S region is $L$. We consider $s$-wave superconductivity which is induced locally by the proximity effect \cite{LF08PRL,Stanescu10PRB,DZhang11PRB,Maier12PRL}.
Then, the junction can be described by two decoupled sets of Bogoliubov-de
Gennes (BdG) equations. The one acting on the basis
$\Psi({\bf r})=(\psi_{+,\uparrow},\psi_{+,\downarrow},\psi_{-,\downarrow}^{\dagger},-\psi_{-,\uparrow}^{\dagger})^{T}$
reads
\begin{equation}
\begin{pmatrix}H(\hat{{\bf K}})-\mu({\bf r}) & \Delta({\bf r})\\
\Delta^{*}({\bf r}) & \mathcal{T}H(-\hat{{\bf K}})\mathcal{T}^{-1}+\mu({\bf r})
\end{pmatrix}\Psi({\bf r})=E\Psi({\bf r}),\label{eq:BdG_Hamiltonian}
\end{equation}
where the gate-voltage-tunable chemical potential $\mu({\bf r})$
is assumed to vary stepwise, $\mu({\bf r})$$=$$\mu_{S}$ in S ($|y|\leq L/2$) and $\mu({\bf r})$$=$$\mu_{L/R}$ in the left/right QH
region ($|y|>L/2$). $\Delta({\bf r})=\Delta_{0}\Theta(L/2-|y|)$
with $\Theta(y)$ the Heaviside function is the pairing
potential. It couples electron and hole excitations from different
Dirac cones of time-reversal partners. The magnetic fields are taken into account via the vector potential $\bm{A}(y)=-yB_{L}\Theta(-y-L/2)\hat{x}-yB_{R}\Theta(y-L/2)\hat{x}$
and substitute the wavevector operators as $\hat{{\bf K}}=\hat{{\bf k}}-e{\bf A}({\bf r})/\hbar$ \cite{Note1},
where $e<0$ is the elementary charge. The other BdG equation acting
on the basis $(\psi_{-,\uparrow},\psi_{-,\downarrow},\psi_{+,\downarrow}^{\dagger},-\psi_{+,\uparrow}^{\dagger})^{T}$
takes the same form as Eq.\ (\ref{eq:BdG_Hamiltonian}) but replacing
$\kappa$ by $-\kappa$.

\textit{Scattering approach.}---In the QH-S-QH junction,
the simple way of wavefunction matching to study transport \cite{Blonder1982}
is not longer applicable. Hence, we need to generalize the scattering
approach \cite{Tamura91PRB} for Dirac fermions under strong magnetic
fields. We assume hard-wall boundary conditions
in $\hat{x}$ direction. In the QH regions, we expand the electron
wavefunction by a complete set of quantum-well states
\begin{equation}
\Psi^{\Lambda}({\bf r})=e^{ik_{y}y-ieB_{\Lambda}xy/\hbar}\sum_{j=1}^{\infty}\chi_{j}(x)f_{j},\label{eq:expansion-Dirac}
\end{equation}
where $\chi_{j}(x)$$=$$\sqrt{2/W}\sin\left[j\pi\left(x/W+1/2\right)\right]$
for $|x|\leq W/2,$ and $0$ otherwise. $f_{j}$ are spinors with only electron
components in the basis used in Eq.~(\ref{eq:BdG_Hamiltonian}).
The phase factor $e^{-ieB_{\Lambda}xy/\hbar}$ stems from the gauge
transformation from ${\bf A}=xB_{\Lambda}\hat{y}$ to $-yB_{\Lambda}\hat{x}$.
$\Lambda\in\{L,R\}$ distinguishes the left and right QH regions.
Plugging Eq.~(\ref{eq:expansion-Dirac}) into the BdG equation, making use of
the inner products between the $\chi_{j}(x)$ and then solving
an eigen equation \cite{SupplementaryMaterial}, we obtain the allowed $k_{y}$ and $f_{j}$ for a given $E$.
The real solutions of $k_{y}$ correspond to propagating channels. All the real $k_y$ together form the excitation energy spectrum of (quasi-)particles.
Similarly, we find the basis wavefunction for holes taking into account the phase factor $e^{ieB_{\Lambda}xy/\hbar}$
and only hole components in $f_{j}$. In S, the
electron and hole components are mixed and the wavefunction is also
expanded in terms of $\chi_{j}(x)$ but without a phase
factor.

With the solutions of wavefunction in each individual region, we construct
the scattering states in the junction. The expansion coefficients which measure
scattering amplitudes between incident and outgoing channels are found by matching the wavefunction of the scattering state and
its derivative at the QH-S interfaces. Summing the squared absolute values of the corresponding scattering amplitudes associated with propagating channels and normalized by the channel velocities, we obtain finally the probabilities of normal
reflection (NR) $R_{ee}$, LAR $R_{eh}$, EC $T_{ee}$, and CAR $T_{eh}$, respectively \cite{SupplementaryMaterial}.

\textit{Landau level spectra in the QH regions.}---We analyze the energy spectra in the QH regions, which can provide
helpful insights into the transport properties of the junction
and the search for perfect CAR. In the QH regions, the energy spectra
for electrons and holes are decoupled and formed by a series of discrete
Landau levels. The guiding centers of electron and hole wavefunctions
are determined by $x=\pm \hbar k_{y} /eB$, respectively. The Landau
levels are flat when they are away from the edges at $x=\pm W/2$.
In the limit $\kappa\ll v \sqrt{\hbar/|eB|}$, which corresponds to a small quadratic
term in Eq.~(\ref{eq:originalModel}), the Landau levels in
the bulk are given by\begin{subequations}\label{eq:LandauLevel}
\begin{eqnarray}
E_{\nu\pm} & = & \pm v\sqrt{2\nu |eB|/\hbar},\ \ \ \ \nu=1,2,\cdots,\label{eq:LLa}\\
E_{0} & = & 0.\label{eq:LLb}
\end{eqnarray}
\end{subequations}The energies are measured from the Fermi level $\mu$.
These Landau levels can be found alternatively exploiting ladder
operators \cite{ZhangSB14PRB,ZhangSB14srep}. In contrast, when close to the edges,
all Landau levels exhibit finite dispersion. The positive levels $E_{\nu+}$
bend upward while the negative levels $E_{\nu-}$ bend downward when
approaching the edges. This behavior implies that electrons and holes move
only in chiral channels close to the edges with their velocities given
by $dE_{v\pm/0}/dk_{y}$. Interestingly, both the zeroth Landau
levels of electrons and holes $E_{0}$, which are particular for the Dirac
fermions, bend either upward or downward, depending on the magnetic field direction. Moreover, they have the same spin polarization,
as indicated by the arrows in Figs.~\ref{fig:schematic-nSp}(c-f),
whose direction also depends on the sign of the magnetic field. For $\mu=0$, the Landau level spectra of electrons
and holes coincide. A finite $\mu$, however, shifts
the spectra oppositely by $\mp\mu$.
As a result, in the quantum limit $0$$<$$|\mu|$$<$$v \sqrt{2|eB|/\hbar}$ \citep{Note3}, only
a single chiral electron channel is maintained at the Fermi level
and contributes to transport in one BdG block, whereas only a single
chiral hole channel with opposite spin polarization contributes in
the other block. Note that the two BdG blocks are completely decoupled in our system.
The remarkable single-channel mechanism is unique to this hybrid junction
which is time-reversal symmetric in the absence of magnetic fields.

\begin{figure}[h]
\centering

\includegraphics[width=8.2cm]{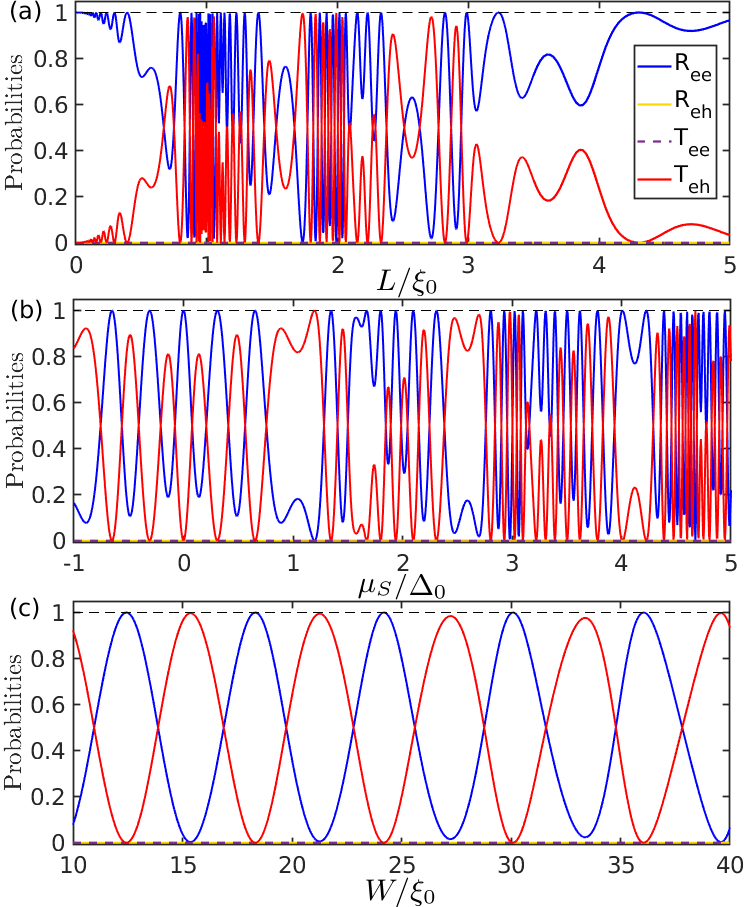}

\caption{Zero-energy probabilities of NR $R_{ee}$ (blue), LAR $R_{eh}$ (yellow), EC $T_{ee}$ (purple)
and CAR $T_{eh}$ (red) as functions of (a) the length $L$,
(b) chemical potential $\mu_{S}$, or (c) junction width $W$.
Here, $B=32B_0$ and other parameters for each panel are the same as in Fig.\ \ref{PerfectCAR_nSp(B)}. }

\label{PerfectCAR_nSp}
\end{figure}
\textit{Perfect CAR.}---The
single-channel mechanism in the quantum limit (described above) is realized in two distinct setups of the QH-S-QH junction, namely,
the p-S-n junction with $\mu_{R}\mu_{L}$$<$$0$ and $B_{L}B_{R}$$>$$0$ and
the n-S-n junction with $\mu_{R}\mu_{L}$$>$$0$ and $B_{L}B_{R}$$<$$0$. In
these setups, for a given BdG block, only electron
channels are allowed in one QH region whereas only hole channels in
the other one. Thus, the processes of EC and LAR
are completely suppressed. We are left with NR and CAR. If an electron stems
from one Cooper pair in S and goes to one QH lead, then the other electron from the pair must
go to the other lead. Note that, these setups are the counterparts
to junctions formed by helical liquids, where NR and
CAR are forbidden \cite{Adroguer10PRB,Crepin15PRB}.
As the chiral edge channels are topologically protected and prohibit backscattering, we expect the setups to be robust against weak disorder and non-ideal conditions regarding interfaces and potential variations \citep{Note4}.
The two setups share many intriguing properties concerning CAR, which we discuss in the following.

We take the p-S-n junction for illustration. For definiteness but
without loss of generality, we assume a positive magnetic field $B_{L}$$=$$B_{R}$$\equiv $$B$$>$$0$
and negative/positive chemical potential in the left/right QH region.
Then, a single chiral electron
channel exists in the left QH region while a single anti-chiral hole
channel in the right QH region. Thus, NR and CAR are
cross-edge processes, i,e., the incident channels and reflected channels
are at the different strip edges, see Figs.\ \ref{fig:schematic-nSp}(a,~b). To explore perfect CAR with $T_{eh}=1$,
we calculate and present in Fig.\ \ref{PerfectCAR_nSp}(a) the zero-energy
probabilities of the four processes as functions of the length $L$. Here, we use $\Delta_0$, $\xi_0\equiv v/\Delta_{0}$ and $B_0\equiv|\hbar/e\xi_0^2|$ as the units for energy, length and magnetic field, respectively. For the typical values $v=100$ meV$\cdot$nm (i.e., $v_F$$\equiv$$v/\hbar$$=$$1.5$$\times$$10^5$ m/s)  and $\Delta_0=1$ meV \cite{Knez12PRL,Wang12science,Wiedenmann16ncomms}, we have $\xi_0=100$ nm and $B_0=0.066$ T which are in an experimentally accessible regime.
In the limit $L\rightarrow0$, the system recovers an n-p junction,
and electrons cannot be converted to holes due to the absence of superconductivity.
In the opposite limit $L\gg\xi_0$, the tunneling
across S is exponentially suppressed. Thus, $T_{eh}=0$ as well. However,
at intermediate length scales, we can find a large or even perfect
CAR. Interestingly, the large CAR persists in the junction
where the length $L$ is longer than the superconducting coherence
length $\xi_0$.

\begin{figure}[htp]
\centering

\includegraphics[width=8.2cm]{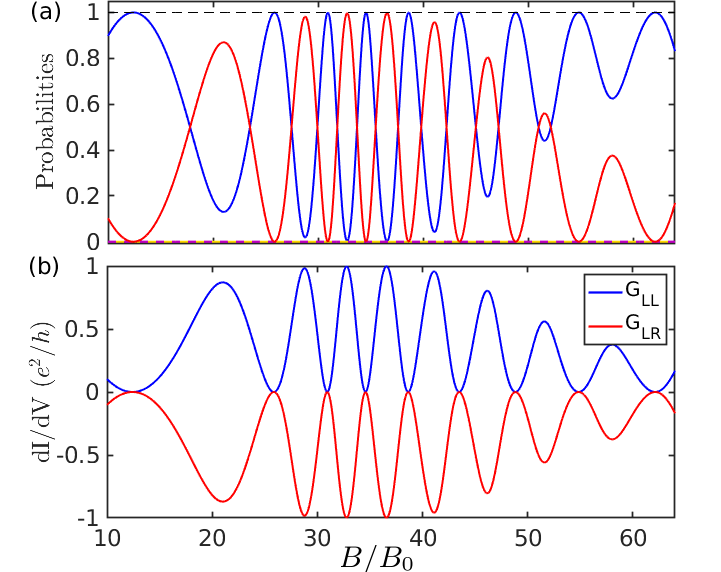}

\caption{(a) Zero-energy probabilities as functions of
$B$ in the quantum limit. Legend is the same as in Fig.\ \ref{PerfectCAR_nSp}(a).
(b) Zero-bias local $G_{\text{LL}}$ and nonlocal differential conductances $G_{\text{LR}}$ as
functions of $B$. We choose $L=2\xi_0$, $W=20\xi_0$, $\mu_{S}=\mu_{R}=-\mu_{L}=3\Delta_0$ and $\kappa=0.01 v\xi_0$.}

\label{PerfectCAR_nSp(B)}
\end{figure}

We also observe Fabry-P$\acute{\text{e}}$rot oscillations with varying
$L$, which stem from the interference effect in S with a finite $\mu_{S}$.
Fabry-P$\acute{\text{e}}$rot oscillations also show up with respect
to the doping $\mu_{S}$ of S and the magnetic field $B$, see Figs.\ \ref{PerfectCAR_nSp}(b)
and\ \ref{PerfectCAR_nSp(B)}(a). The interference occurs not only along
the junction in $\hat{y}$ direction but also across the strip in $\hat{x}$
direction. Thus, analogous oscillations appear with respect to the
junction width $W$, see Fig.\ \ref{PerfectCAR_nSp}(c).
The pattern of oscillations is, however, more regular because the
interference consists of a single pair of propagating modes in $\hat{x}$
direction, in contrary to the interference in $\hat{y}$ direction
which also involves modes with decaying oscillation behavior. Therefore, we are able to obtain
perfect CAR. It is possible to control CAR by length, doping of S, magnetic
field strength, and junction width. Finally, we stress that the large
CAR occurs at zero energy, which indicates the exempt from
a fine tuning of bias voltage.

Next, we study the transport signature of CAR.
We calculate the local and nonlocal differential conductances, which
are defined as $G_{\text{LL}}\equiv dI_{L}/dV_{L}|_{V_{R}=0}$ and
$G_{\text{LR}}\equiv dI_{R}/dV_{L}|_{V_{R}=0}$, respectively, by
using the extended Blonder-Tinkham-Kapwijk theory \cite{Lambert93JPCM,Anantram96PRB}.
Here, $I_{L/R}$ and $V_{L/R}$ are the measured current and applied
bias voltage in the left/right QH region, respectively. S is grounded.
In the two setups, LAR and EC
are eliminated completely so that NR and CAR dominate
the local and nonlocal transport, respectively. The conductances at zero
temperature are
\begin{align}
G_{\text{LR}}=-G_{\text{LL}} & =-(e^{2}/h)T_{eh}.
\end{align}
Here, $T_{eh}=1-R_{ee}$, as required by the particle conservation,
and the bias voltage enters the conductances as excitation energy
via $T_{eh}$. $G_{\text{LR}}$ is negative
and exactly opposite to $G_{\text{LL}}$, as shown in
Fig.\ \ref{PerfectCAR_nSp(B)}(b). Importantly, $G_{\text{LR}}$
provides not only a transport signature but also a direct measurement
of CAR.

\textit{Spin injection in the p-S-n junction.}---The
difference of the two setups manifests mainly in the spin injection
into S, which we now clarify. In the n-S-n junction, the incident
electron and reflected hole carry opposite spin, see Figs.\ \ref{fig:schematic-nSp}(e,f).
This implies that two electrons with opposite spin are absorbed into S to form a spin-singlet Cooper pair, as we expect for $s$-wave
superconductivity. Therefore, we have no spin transport between S
and the QH regions.

\begin{figure}[h]
\centering

\includegraphics[width=8.5cm]{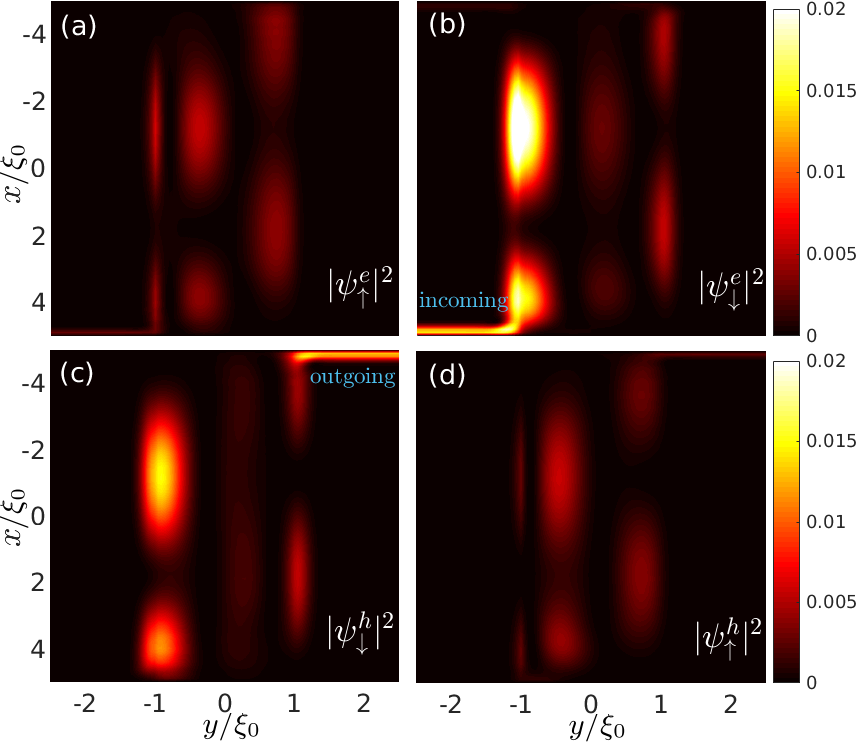}

\caption{Contour plots of the densities of (a) spin-up, (b) spin-down electrons,
(c) spin-down and (d) spin-up holes of a zero-energy scattering state in the p-S-n junction. Here, $B=32B_0$, $W=10\xi_0$, $L=2\xi_0$ and other parameters
are the same as in Fig.\ \ref{PerfectCAR_nSp(B)}.}

\label{fig:map}
\end{figure}

However, this is not the case for the p-S-n junction. The reflected
hole carries spin down which is remarkably the same as that carried
by the incident electron, see Figs.\ \ref{fig:schematic-nSp}(c,d). To further confirm this, we calculate the density distributions of the four
components of a scattering state near the junction in Fig.\ \ref{fig:map}.
In S, the four components mix together due the presence of superconductivity
and strong spin-orbit coupling, and they oscillate in both $\hat{x}$
and $\hat{y}$ directions, reflecting the aforementioned interference
effect. We see clearly that the incident electron
carrying spin down at the lower edge is converted through S
as a hole carrying also spin down at the upper edge into the other
region. Therefore, we have an equal-spin CAR which effectively
pumps two equal spins into S. The equal-spin CAR implies the creation of equal-spin triplet Cooper pairs in S \cite{Benjamin06PRB,Breunig18PRL,Fleckenstein18PRB,Note2}, which are of interest in superconducting spintronics \cite{Linder15nphys}.
Following the approach of Ref.\ \cite{Breunig18PRL}, we predict the
value of spin pumped into S explicitly as
\begin{equation}
\bar{S}_{z}=-(h/2\pi)T_{eh}.
\end{equation}
The spin injection is purely contributed by CAR. We have a quantized
spin injection of $-h/2\pi$ per CAR event. For perfect CAR, we
obtain a perfect spin injection.
%If we change the signs of both $\mu_{L}$ and $\mu_{R}$ or $B_L$ and $B_R$, then we pump instead $h/2\pi$ into S.

\textit{Summary.}---We have developed a scattering approach to investigate a 2D Dirac QH-S-QH junction. We have proposed two different setups, which
exploit the particular properties of the zeroth Landau levels of the
Dirac fermions in the quantum limit, for realizing high-efficiency
and controllable CAR without fine tuning of bias voltage. The differential conductances provide a direct measurement of CAR. We have identified a quantized spin injection in the p-S-n junction.

\begin{acknowledgements}
We thank C. Gould, L. W. Molenkamp, M. Stehno, and G. Tang for helpful discussions.
This work was supported by the DFG (SPP1666,  SFB1170 ``ToCoTronics"), the W\"urzburg-Dresden Cluster of Excellence ct.qmat, EXC2147, project-id 39085490, and the Elitenetzwerk Bayern Graduate School on ``Topological
Insulators".
\end{acknowledgements}

%\bibliographystyle{apsrev4-1}
%\bibliography{ReferenceSB}


\onecolumngrid

\appendix
\clearpage
\newpage
\renewcommand{\thepage}{S\arabic{page}}
\renewcommand{\thefigure}{S\arabic{figure}}

\begin{center}
\bf{\large Supplemental Material}
\end{center}

In this Supplemental Material, we present the calculations of the
excitation energy spectrum in each region and the scattering probabilities
in the QH-S-QH junctions. \ \
\newline
\newline

\twocolumngrid
\section{}

We start with the QH regions. The electrons and holes are decoupled.
Thus, we calculate them separately. In the presence of a magnetic
field $B$ perpendicular to the strip plane, the Dirac equation for
electrons reads
\begin{align}
(E+\mu)\psi_{e}({\bf r})= & \{[\kappa(x\xi_{B}-i\partial_{y})^{2}-\kappa\partial_{x}^{2}]s_{z}\nonumber \\
 & +(x\xi_{B}-i\partial_{y})s_{y}-i\partial_{x}s_{x}\}\psi_{e}({\bf r}),\label{eq:Dirac-equation}
\end{align}
where the gauge for the magnetic field ${\bf A}=xB\hat{y}$ is adopted
and $\xi_{B}\equiv-eB/\hbar$. We assume hard-wall boundary conditions
for the strip edges and expand the wavefunction in terms of quantum-well
states
\begin{equation}
\psi_{e}({\bf r})=e^{ik_{y}y}f(x),\ \ f(x)=\sum_{j=1}^{\infty}\chi_{j}(x)f_{j},\label{eq:expansion-Dirac}
\end{equation}
where $f_{j}$ are two component spinors. The basis functions $\chi_{j}(x)$
are given by
\begin{equation}
\chi_{j}(x)=\begin{cases}
\sqrt{2/W}\sin\left[j\pi\left(x/W+1/2\right)\right], & |x|\leq W/2,\\
0, & |x|>W/2.
\end{cases}
\end{equation}
Plugging Eq.\ (\ref{eq:expansion-Dirac}) into Eq.\ (\ref{eq:Dirac-equation}),
this leads to
\begin{align}
0= & \{\kappa k_{y}^{2}s_{z}+2\kappa xk_{y}\xi_{B}s_{z}+\kappa x^{2}\xi_{B}^{2}s_{z}+\kappa(j\pi/W)^{2}s_{z}\nonumber \\
 & +k_{y}s_{y}+x\xi_{B}s_{y}-E-\mu\}\sideset{}{_{j}}\sum\chi_{j}(x)f_{j}\nonumber \\
 & -is_{x}\sideset{}{_{j}}\sum\partial_{x}\chi_{j}(x)f_{j},
\end{align}
Multiplying both sides by $\chi_{t}(x)$ and integrating over $x$,
we obtain
\begin{align}
0= & \sideset{}{_{j}}\sum\{[\kappa k_{y}^{2}s_{z}+\kappa(j\pi/W)^{2}s_{z}+k_{y}s_{y}-E-\mu]\delta_{t,j}\nonumber \\
 & \ \ \ \ \ +2\kappa k_{y}\xi_{B}s_{z}\left\langle x\right\rangle _{t,j}+\kappa\xi_{B}^{2}s_{z}\left\langle x^{2}\right\rangle _{t,j}\nonumber \\
 & \ \ \ \ \ +\xi_{B}s_{y}\left\langle x\right\rangle _{t,j}-is_{x}\left\langle \partial_{x}\right\rangle _{t,j}\}f_{j},\label{eq:eigenEq}
\end{align}
where the inner products between the quantum-well states are given
by
\begin{align}
\left\langle x\right\rangle _{t,j}= & \begin{cases}
0,{\color{white}\text{\ensuremath{\dfrac{}{}}}} & j=t,\\
4tjW[(-1)^{j+t}-1]/[\pi^{2}(t^{2}-j^{2})^{2}], & j\neq t,
\end{cases} \notag \\
\left\langle x^{2}\right\rangle _{t,j}= & \begin{cases}
W^{2}[1-6/(\pi^{2}t^{2})]/12,{\color{white}\text{\ensuremath{\dfrac{}{}}}} & j=t,\\
4W^{2}tj[(-1)^{j+t}+1]/[\pi^{2}(t^{2}-j^{2})^{2}], & j\neq t,
\end{cases} \notag\\
\left\langle \partial_{x}\right\rangle _{t,j}= & \begin{cases}
0,{\color{white}\text{\ensuremath{\dfrac{}{}}}} & j=t,\\
2tj[(-1)^{j+t}-1]/[W(j^{2}-t^{2})], & j\neq t.
\end{cases}
\end{align}
Choosing a large number $N$ of basis functions, we recast the equation
in a $4N$-dimensional matrix form
\begin{equation}
\begin{pmatrix}0 & 1\\
\hat{T} & \hat{U}
\end{pmatrix}\begin{pmatrix}f\\
k_{y}f
\end{pmatrix}=k_{y}\begin{pmatrix}f\\
k_{y}f
\end{pmatrix},\label{eq:ExpandCoefficient-2}
\end{equation}
where $\hat{M}$ indicates that $M$ is a matrix. The elements of
the relevant matrices are given by
\begin{align}
T_{t,j}= & [(E+\mu)s_{z}/\kappa-(j\pi/W)^{2}]\delta_{t,j}-\xi_{B}^{2}\left\langle x^{2}\right\rangle _{t,j}\nonumber \\
 & +is_{x}\xi_{B}\left\langle x\right\rangle _{t,j}/\kappa-s_{y}\left\langle \partial_{x}\right\rangle _{t,j}/\kappa, \notag \\
U_{t,j}= & is_{x}\delta_{t,j}/\kappa-2\xi_{B}\left\langle x\right\rangle _{t,j}.
\end{align}
Diagonalizing the $4N\times4N$ matrix in Eq.\ (\ref{eq:ExpandCoefficient-2})
for a given energy $E$, we can find $2N$ pairs of eigenvalues $\pm k_{j}^{e}$
for the wavenumbers and the corresponding $4N$ spinors $f_{j}$ (corresponding to positive eigenvalues $+ k_{j}^{e}$) and
$g_{j}$ (corresponding to negative eigenvalues $-k_{j}^{e}$). The real solutions of $k_{j}^{e}$ correspond to propagating
channels. All real $k_{j}^{e}$ together form the electron spectrum.

The Dirac equation for holes is given by
\begin{align}
(\mu-E)\psi_{h}({\bf r})= & \{[\kappa(x\xi_{B}+i\partial_{y}){}^{2}-\kappa\partial_{x}^{2}]s_{z} \label{eq:Dirac-equation-1} \\
 & +(-x\xi_{B}-i\partial_{y})s_{y}-i\partial_{x}s_{x}\}\psi_{h}({\bf r}).  \nonumber
\end{align}
 Similarly, we expand the wavefunction as
\begin{equation}
\psi_{h}({\bf r})=e^{ik_{y}y}h(x),\ \ h(x)=\sum_{j=1}^{\infty}\chi_{j}(x)h_{j},\label{eq:expansion-Dirac-1}
\end{equation}
and derive the following equation
\begin{equation}
\begin{pmatrix}0 & 1\\
\hat{T}' & \hat{U}'
\end{pmatrix}\begin{pmatrix}h\\
|\kappa|^{1/2}k_{y}h
\end{pmatrix}=|\kappa|^{1/2}k_{y}\begin{pmatrix}h\\
|\kappa|^{1/2}k_{y}h
\end{pmatrix},\label{eq:ExpandCoefficient-2-1}
\end{equation}
where
\begin{align}
T_{t,j}'= & \left[(-E+\mu)s_{z}/\kappa-(j\pi/W)^{2}\right]\delta_{t,j}-\xi_{B}^{2}\left\langle x^{2}\right\rangle _{t,j}\nonumber \\
 & -i\xi_{B}s_{x}\left\langle x\right\rangle _{t,j}/\kappa-s_{y}\left\langle \partial_{x}\right\rangle _{t,j}/\kappa,\notag \\
U_{t,j}'= & is_{x}\delta_{t,j}/\kappa+2\xi_{B}\left\langle x\right\rangle _{t,j}.
\end{align}

\begin{figure}[h]
\centering

\includegraphics[width=7.5cm]{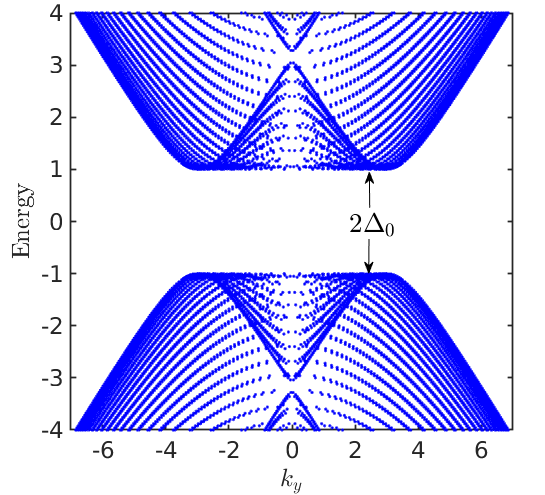}

\caption{Excitation energy spectrum (in units of $\Delta_{0}$) of the superconductor.
$k_{y}$ is in units of $1/\xi_{0}$. The superconductor is full gaped.
Here, $W=20\xi_{0}$, $\mu_{S}=3\Delta_{0}$ and $\kappa=0.01v\xi_{0}$. }

\label{fig:SC-sepctrum}
\end{figure}

In the superconductor, the wavefunction is taken as
\begin{equation}
\psi({\bf r})=e^{iq_{y}y}\varphi(x),\ \ \varphi(x)=\sum_{j=1}^{\infty}\chi_{j}(x)\varphi_{j},\label{eq:expansion-BdG}
\end{equation}
where $\varphi_{j}$ are four-component spinors in Nambu space. Then,
we have the BdG equation as
\begin{align}
0= & [(-\kappa\partial_{x}^{2}+\kappa q_{y}^{2})\tau_{z}s_{z}-\mu\tau_{z}-E\nonumber \\
 & +q_{y}\tau_{z}s_{y}-i\tau_{z}s_{x}\partial_{x}+\Delta_{0}\tau_{x}]\sideset{}{_{j}}\sum\chi_{j}(x)\varphi_{j}.\label{eq:Dirac-QWE-1}
\end{align}
Next, we multiply Eq.\ (\ref{eq:Dirac-QWE-1}) by $\chi_{t}(x)$
and integrate over $x$. The BdG equation becomes
\begin{equation}
\begin{pmatrix}0 & 1\\
\hat{S} & \hat{P}
\end{pmatrix}\begin{pmatrix}\varphi\\
q_{y}\varphi
\end{pmatrix}=q_{y}\begin{pmatrix}\varphi\\
|q_{y}\varphi
\end{pmatrix},\label{eq:ExpandCoefficient-S}
\end{equation}
where
\begin{align}
S_{t,j}= & [E\tau_{z}s_{z}+\mu s_{z}-\kappa(j\pi/W)^{2}\nonumber \\
 & -i\Delta_{0}\tau_{y}s_{z}]\delta_{t,j}/\kappa-s_{y}\left\langle \partial_{x}\right\rangle _{t,j}/\kappa, \notag\\
P_{t,j}= & is_{x}\delta_{t,j}/\kappa.
\end{align}
Similar to the QH regions, we find $8N$ wavenumbers $q_{y}=q_{j}$
and $8N$ spinors $\varphi_{j}$ from the BdG Eq.\ (\ref{eq:ExpandCoefficient-S}).
The energy spectrum of the superconductor is given by the real $q_{j}$
for all $E$. A typical spectrum is presented in Fig.\ \ref{fig:SC-sepctrum}.
The superconductor has a full gap of $2\Delta_{0}$, as expected by
time-reversal symmetry.

With the wavefunction in each individual region, we now calculate
the scattering probabilities. We denote $f_{l}^{\Lambda}(x)$ the
$N$ wavefunctions for electron modes with $k_{l,\Lambda}^{e}$ and
positive velocities, while $g_{l,\Lambda}(x)$ the $N$ wavefunctions
for electron modes with $-k_{l,\Lambda}^{e}$ and negative velocities.
Similarly, we have $\rho_{l,\Lambda}(x)$ and $h_{l,\Lambda}(x)$
for the right- and left-moving hole modes with wavenumbers $\pm k_{l,\Lambda}^{h}$,
respectively. Here, $\Lambda\in\{L,R\}$ distinguishes the left and
right QH regions. Then, the wavefunction of a scattering state of
an electron with the wavenumber $k_{n}^{e}$ incident from the left
QH region to the superconductor is built up as
\begin{align}
\Psi_{n}=\begin{cases}
\sideset{}{_{l}}\sum\Bigg[e^{ieB_{L}xy/\hbar}\delta_{l,n}e^{ik_{n,L}^{e}y}\begin{pmatrix}f_{n}^{L}(x)\\
0
\end{pmatrix}\\
+e^{{\color{red}ieB_{L}xy}/\hbar}B_{l,n}e^{-ik_{l,L}^{e}y}\begin{pmatrix}g_{l}^{L}(x)\\
0
\end{pmatrix}\\
+e^{-ieB_{L}xy/\hbar}A_{l,n}e^{-ik_{l,L}^{h}y}\begin{pmatrix}0\\
h_{l}^{L}(x)
\end{pmatrix}\Bigg], & y<0,\\
\sideset{}{_{l}}\sum\alpha_{l,n}e^{iq_{l}y}\varphi_{l}(x){\color{white}\text{\ensuremath{\dfrac{{\color{white}\text{\ensuremath{\dfrac{}{}}}}}{{\color{white}\text{\ensuremath{\dfrac{}{}}}}}}}}, & 0\leqslant y\leqslant L,\\
\sideset{}{_{l}}\sum\Bigg[e^{ieB_{R}xy/\hbar}C_{l,n}e^{ik_{l,R}^{e}y}\begin{pmatrix}f_{l}^{R}(x)\\
0
\end{pmatrix}\\
+e^{-ieB_{R}xy/\hbar}D_{l,n}e^{ik_{l,R}^{h}y}\begin{pmatrix}0\\
\rho_{l}^{R}(x)
\end{pmatrix}\Bigg], & y>L,
\end{cases}
\end{align}
where the matrix elements $A_{l,n}$, $B_{l,n}$, $C_{l,n}$ and $D_{l,n}$
represent the scattering amplitudes of local Andreev reflection, normal
reflection, electron co-tunneling and crossed Andreev reflection,
respectively. The phase factors $e^{\pm ieB_{\Lambda}xy/\hbar}$ stem
from the gauge transformation of ${\bf A}=xB_{\Lambda}\hat{y}$ to
$-yB_{\Lambda}\hat{x}$ for the transport problem. The wavefunction
$\Psi_{n}({\bf r})$ and its derivative are continuous at the two
QH-S interfaces at $y=\pm L/2$, which yield the following equation
to find the scattering amplitudes

\begin{widetext}
\begin{equation}
\begin{pmatrix}\begin{array}{c}
0\\
-\hat{h}^{L}
\end{array} & \begin{array}{c}
-\hat{g}^{L}\\
0
\end{array} & \begin{array}{c}
0\\
0
\end{array} & \begin{array}{c}
0\\
0
\end{array} & \hat{\varphi}\\
\begin{array}{c}
0\\
\hat{\Gamma}^{L}\hat{h}^{L}+\hat{h}^{L}\hat{k}_{L}^{h}
\end{array} & \begin{array}{c}
\hat{g}^{L}\hat{k}_{L}^{e}-\hat{\Gamma}^{L}\hat{g}^{L}\\
0
\end{array} & \begin{array}{c}
0\\
0
\end{array} & \begin{array}{c}
0\\
0
\end{array} & \hat{\Phi}\\
\begin{array}{c}
0\\
0
\end{array} & \begin{array}{c}
0\\
0
\end{array} & \begin{array}{c}
-\hat{f}^{R}e^{i\hat{k}_{R}^{e}L}\\
0
\end{array} & \begin{array}{c}
0\\
-\hat{\rho}^{R}e^{-i\hat{k}_{R}^{h}L}
\end{array} & \hat{\varphi}e^{i\hat{q}L}\\
\begin{array}{c}
0\\
0
\end{array} & \begin{array}{c}
0\\
0
\end{array} & \begin{array}{c}
(-\hat{\Gamma}^{R}\hat{f}^{R}-\hat{f}^{R}\hat{k}_{R}^{e})e^{i\hat{k}_{R}^{e}L}\\
0
\end{array} & \begin{array}{c}
0\\
(\hat{\Gamma}^{R}\hat{h}^{R}-\hat{\rho}^{R}\hat{k}_{R}^{h})e^{-i\hat{k}_{R}^{h}L}
\end{array} & \hat{\Phi}e^{i\hat{q}L}
\end{pmatrix}\begin{pmatrix}\hat{A}\\
\hat{B}\\
\hat{C}\\
\hat{D}\\
\\
\hat{\alpha}\\
\\
\end{pmatrix}=\begin{pmatrix}\hat{f}^L\\
0\\
\hat{f}^L\hat{k}_{L}^{e}+\hat{\Gamma}\hat{f}^L\\
0\\
0\\
0\\
0\\
0
\end{pmatrix},\label{eq:final-eq}
\end{equation}
\end{widetext}where $\Gamma_{t,j}^{\Lambda}=-eB_{\Lambda}\left\langle x\right\rangle _{t,j}/\hbar$
and $\Phi_{t,j}=\varphi_{t,j}q_{j}$. The hat $\hat{...}$ again indicates
a matrix. $\hat{f}^{L(R)}$ are $2N\times2N$ matrices formed by the spinors $f_j$ associated with positive $+k_{j,L(R)}^e$ in the left (right) lead. $\hat{g}^{L(R)}$ are $2N\times2N$ matrices formed by the spinors $g_j$ associated with positive $-k_{j,L(R)}^e$ in the left (right) lead. $\hat{\rho}^{L(R)}$ and $\hat{h}^{L(R)}$ are similar to  $\hat{g}^{L(R)}$ and $\hat{f}^{L(R)}$ but for the hole counterplart.  $ \hat{\varphi}$ and $\hat{\Phi}$ are $4N\times8N$ matrices
and $\hat{\alpha}$ is a $8N\times 2N$ matrix. On deriving Eq.\ (\ref{eq:final-eq}),
we have made use of the inner products of the quantum-well states.
Solving Eq.\ (\ref{eq:final-eq}), we find the matrices $\hat{A},$
$\hat{B},$ $\hat{C},$ and $\hat{D}$. Then, the probabilities (multiplying
the number of available incident channels) of the four scattering
processes, normal reflection $R_{ee},$ local Andreev reflection $R_{eh},$
electron co-tunneling $T_{ee},$ and crossed Andreev reflection $T_{eh}$,
are calculated explicitly as
\begin{figure}[t]
\centering

\includegraphics[width=8.5cm]{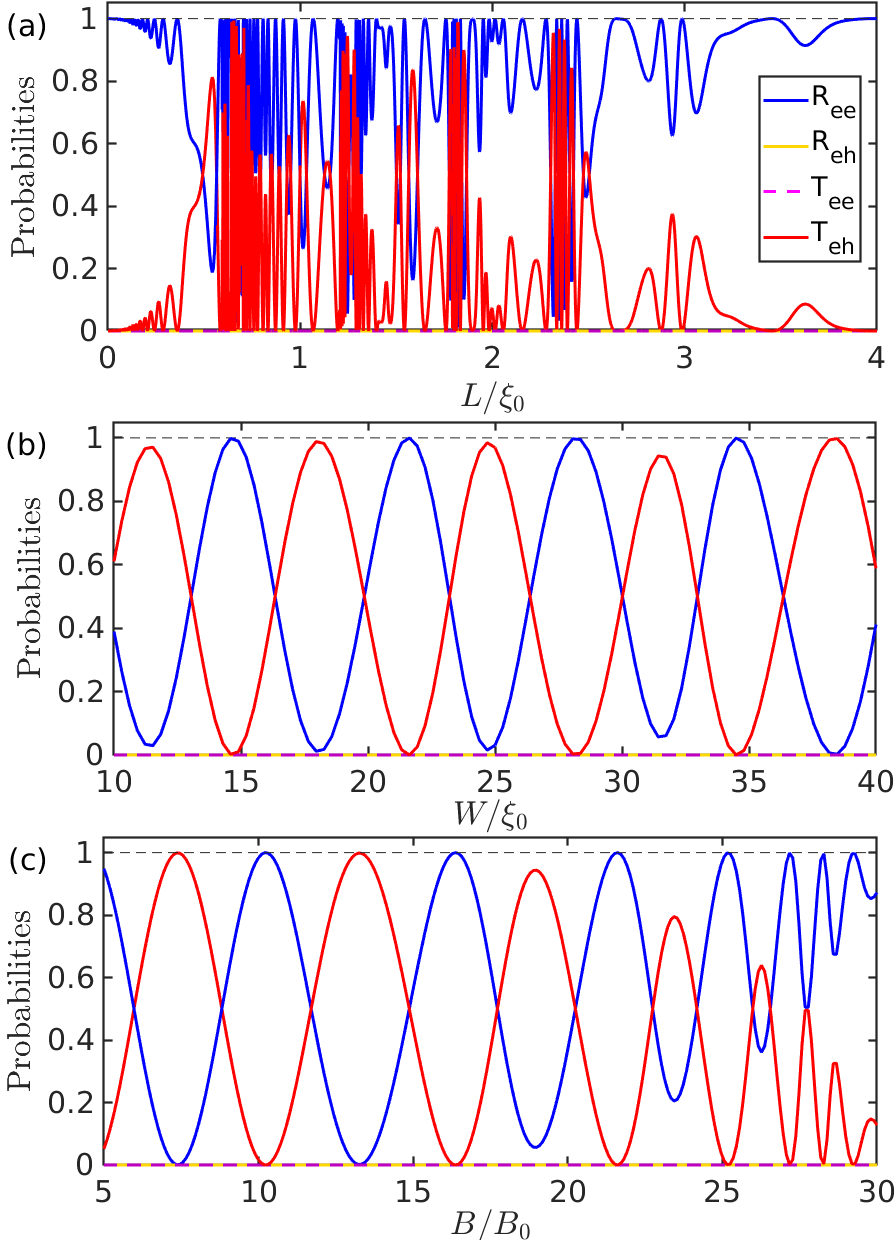}

\caption{(a) Zero-energy probabilities of $R_{ee}$ (blue), $R_{eh}$ (yellow),
$T_{ee}$ (purple) and $T_{eh}$ (red) in the p-S-n junction as functions
of the length $L$ for $W=20\xi_{0}$ and $B=32B_{0}$; (b) Zero-energy
probabilities as functions of the junction width $W$ for $L=2\xi_{0}$
and $B=15B_{0}$; (c) Zero-energy probabilities as functions of of
the magnetic field $B$ for $W=20\xi_{0}$ and $L=2\xi_{0}$. Other
parameters for all panels are $\mu_{S}=5\Delta_{0}$ and $\mu_{L}=-\mu_{R}=-2\Delta_{0}$.}

\label{fig:CAR-pSn}
\end{figure}

\begin{figure}[H]
\centering

\includegraphics[width=8.3cm]{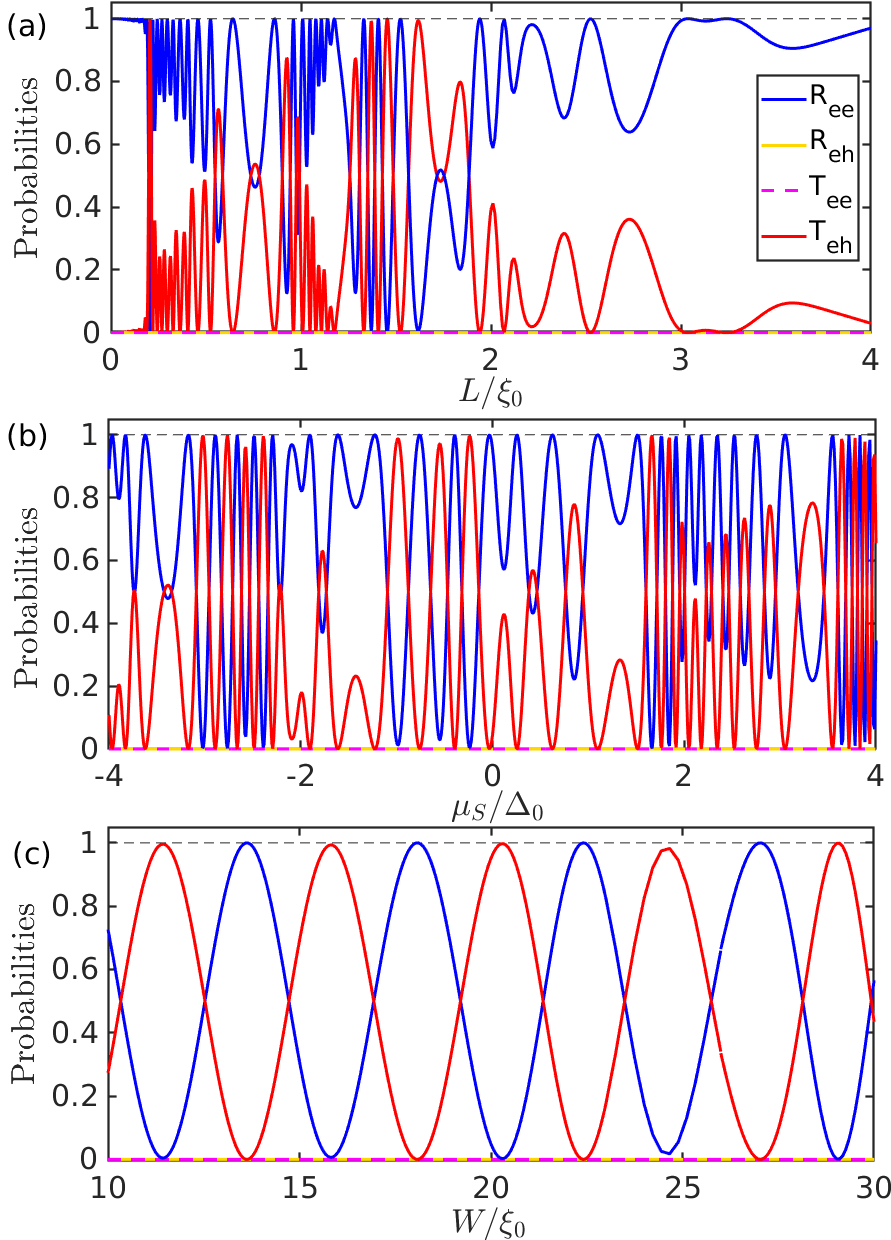}

\caption{(a) Zero-energy probabilities of $R_{ee}$ (blue), $R_{eh}$ (yellow),
$T_{ee}$ (purple) and $T_{eh}$ (red) in the n-S-n junction as functions
of the length $L$ for $W=20\xi_{0}$, $B=32B_{0}$ and $\mu_{S}=-3\Delta_{0}$;
(b) Zero-energy probabilities as functions of of the chemical potential
$\mu_{S}$ for $L=1.5\xi_{0}$, $B=15B_{0}$ and $W=20\xi_{0}$; (c)
Zero-energy probabilities as functions of the junction width $W$
for $L=1.5\xi_{0}$, $B=15B_{0}$ and $\mu_{S}=-3\Delta_{0}$; Other
parameters for all panels are $\mu_{L}=\mu_{R}=-3\Delta_{0}$ and
$\kappa=0.01v\xi_{0}$.}

\label{fig:CAR-nSn}
\end{figure}

\begin{align}
R_{ee} & =\sum_{l,n}|A_{l,n}|^{2}|v_{l,\leftarrow}^{e}/v_{n,\rightarrow}^{e}|, \notag\\
R_{eh} & =\sum_{l,n}|B_{l,n}|^{2}|v_{l,\leftarrow}^{h}/v_{n,\rightarrow}^{e}|,\notag\\
T_{ee} & =\sum_{l,n}|C_{l,n}|^{2}|v_{l,\rightarrow}^{e}/v_{n,\rightarrow}^{e}|,\notag\\
T_{eh} & =\sum_{l,n}|D_{l,n}|^{2}|v_{l,\rightarrow}^{h}/v_{n,\rightarrow}^{e}|,
\end{align}
where the sums run over all available channels with real wavenumbers.
$v_{l,\rightleftarrows}^{\Lambda,e(h)}$ are the velocities in the
$l$th channel in the electron (hole) branch. The arrows indicate
the propagating direction of channels. The velocities are calculated,
respectively, as
\begin{align}
v_{l,\rightarrow}^{\Lambda,e}= & \sum_{j,j'}f_{j,l}^{\Lambda*}\left[s_{y}\delta_{j,j'}+2\kappa s_{z}(\Gamma_{j,j'}^{\Lambda}+k_{l,\Lambda}^{e}\delta_{j,j'})\right]f_{j',l}^{\Lambda}, \notag \\
v_{l,\leftarrow}^{\Lambda,e}= & \sum_{j,j'}g_{j,l}^{\Lambda*}\left[s_{y}\delta_{j,j'}+2\kappa s_{z}(\Gamma_{j,j'}^{\Lambda}-k_{l,\Lambda}^{e}\delta_{j,j'})\right]g_{j',l}^{\Lambda}, \notag\\
v_{l,\rightarrow}^{\Lambda,h}= & \sum_{j,j'}\rho_{j,l}^{\Lambda*}\left[s_{y}\delta_{j,j'}-2\kappa s_{z}(\Gamma_{j,j'}^{\Lambda}-k_{l,\Lambda}^{h}\delta_{j,j'})\right]\rho_{j',l}^{\Lambda}, \notag\\
v_{l,\leftarrow}^{\Lambda,h}= & \sum_{j,j'}h_{j,l}^{\Lambda*}\left[s_{y}\delta_{j,j'}-2\kappa s_{z}(\Gamma_{j,j'}^{\Lambda}-k_{l,\Lambda}^{h}\delta_{j,j'})\right]h_{j',l}^{\Lambda}.
\end{align}

At zero temperature, the local and nonlocal differential conductances
in the sub-gap regime (where $eV<\Delta_{0}$) are given by
\begin{align}
\dfrac{dI_{\text{L}}}{dV_{\text{L}}} & =\dfrac{e^{2}}{h}\left(N_{c}-R_{ee}+R_{eh}\right), \notag\\
\dfrac{dI_{\text{R}}}{dV_{\text{L}}} & =\dfrac{e^{2}}{h}\left(T_{ee}-T_{eh}\right),
\end{align}
respectively, where $N_{c}$ counts the number of incident channels,
and $N_{c}=1$ in the quantum limit. The particle conservation yields
$R_{ee}+R_{eh}+T_{ee}+T_{eh}=N_{c}$.

The main features of the zero-energy probabilities we discussed in
the Letter, namely, 1) the complete suppression of local Andreev reflection
and electron co-tunneling; 2) the accessibility of perfect CAR $T_{eh}=1$
by tuning $L$, $W$, $\mu_{S}$ or $B$ (for the latter three kinds
of tuning, an intermediate length $L$ is required); and 3) the oscillation
behaviors due to the interference effects along $\hat{x}$ and $\hat{y}$
directions; are not restricted to specific values of $\mu_{S}$ and
$\mu_{L/R}$. In Fig.\ \ref{fig:CAR-pSn}, we present the typical
results for the case with $\mu_{S}$ significantly different from
$\mu_{L/R}$. For larger $\mu_{S}$, the oscillations are more rapid
as varying the length $L$ or magnetic field $B$, as we can see by
comparing Fig.\ \ref{fig:CAR-pSn} with Fig.\ 2(a) and 3(a) in the
Letter.

\begin{figure}[H]
\centering

\includegraphics[width=9cm]{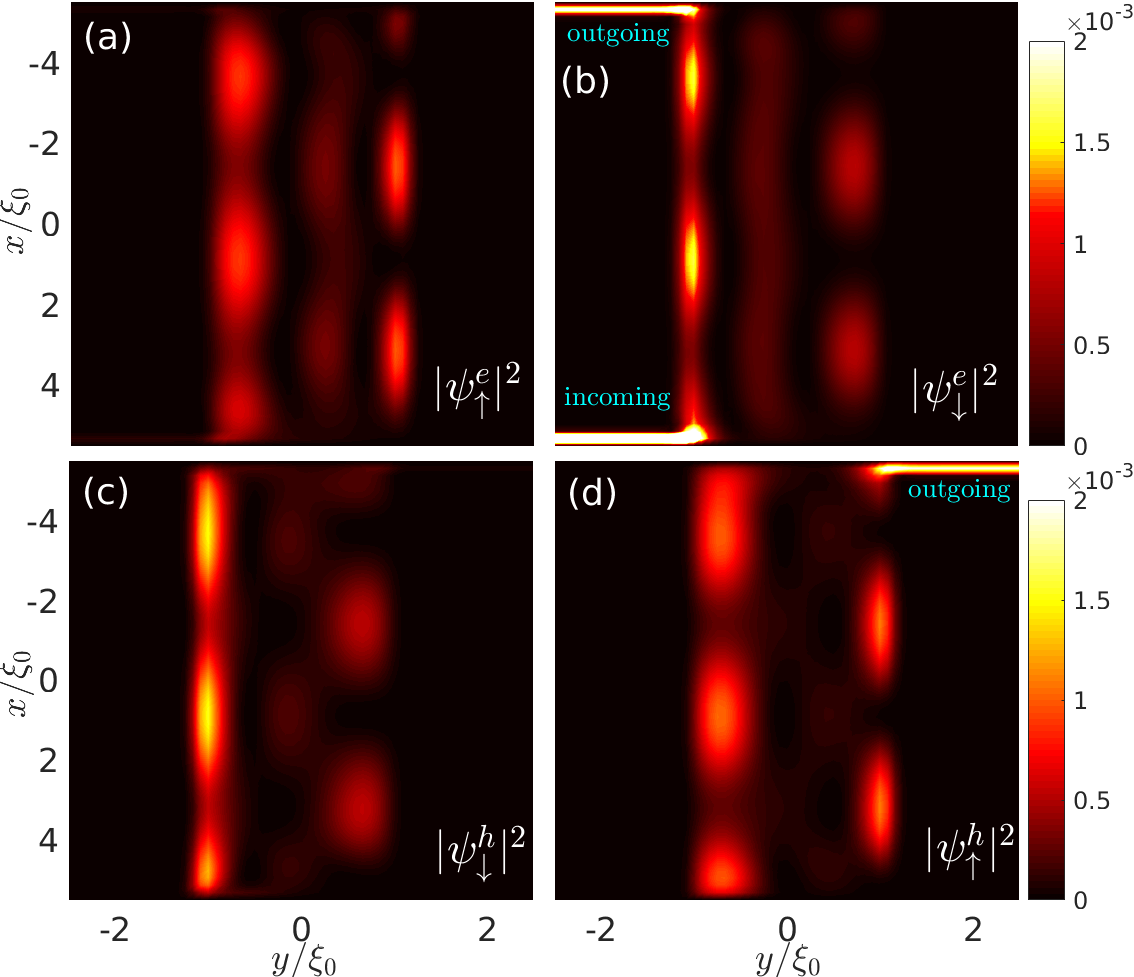}

\caption{Contour plots of the densities of (a) spin-up, (b) spin-down electrons,
(c) spin-down and (d) spin-up holes of a zero-energy scattering state
in the n-S-n junction. Parameters are $B=50B_{0}$, $W=11\xi_{0}$,
$L=2\xi_{0}$ $\mu_{S}=\mu_{L}=\mu_{R}=-3\Delta_{0}$ and $\kappa=0.01v\xi_{0}$.}

\label{fig:map}
\end{figure}

For completeness, we present in Figs.\ \ref{fig:CAR-nSn} and \ref{fig:map}
the typical results of the n-S-n junction. They have the same main
features of those for the p-S-n junction. Although we choose the parameters
$\mu_{S}=\mu_{L}=\mu_{R}=-3\Delta_{0}$ for illustration, the results
are general and not restricted to this choice. As shown in Fig.\ \ref{fig:map},
an incident electron with spin down can be scattered back to the opposite
edge of the same QH lead. It can also be crossed Andreev reflected
into the hole channel with opposite spin in the other lead, which
is different from the case of the p-S-n junction.

\end{document}